\documentclass[twocolumn,secnumarabic,amssymb, nobibnotes, aps, prd]{revtex4-2}
\usepackage{graphicx}
\usepackage{dcolumn}
\usepackage{bm}

\begin{document}

\title{Unified Cranking-Model Description of Bohr and VMI Approache}

\author{Moh’d Abu El Sheikh}
\email[]{mhsr70@gmail.com}
\affiliation{Namangan State Technical University, Namangan, 160100, Uzbekistan}

\author{Abdurahim A. Okhunov,}
\email[]{abdurahimokhun@iium.edu.my}
\homepage[]{www.iium.edu.my}
\affiliation{Department of Science in Engineering, International Islamic University Malaysia, 53100 Kuala Lumpur, Malaysia \\ Department of Physics, Andijan State University, Andijan, 170100, Uzbekistan}

\author{Riad S. Masharfe,}
\affiliation{Department of Physics, Zartqa University, 13132 Zarqa, Jordan}

\author{Anashon A. Yokubbayev}
\affiliation{University of Business and Science, 100160 Namangan, Uzbekistan}

\date{\today}

\begin{abstract}
A unified analysis based on the cranking model is presented, demonstrating that both the Bohr and Variable Moment of Inertia (VMI) models arise as limiting cases of this framework. This result resolves the apparent contradiction between the two models by showing that they correspond to different physical limits of the same formalism. In addition, this analysis explicitly reveals the contribution of Coriolis effects to the rotational energy and moment of inertia.
\end{abstract}


\maketitle

\section{INTRODUCTION}

The Bohr hydrodynamic model ~\cite{BohrMot75} traditionally defines the nuclear moment of inertia as being proportional to the square of the deformation parameter $\left(\Im\propto\beta^2\right)$, where $\beta$ is taken to be independent of the nuclear spin I. However, this assumption of constant $\beta$ has proven inadequate for accurately reproducing the observed spectra of rotating nuclei. Diamond ~\cite{Diamond64} addressed this deficiency by introducing the concept of centrifugal stretching, proposing that $\beta$ is allowed to vary self-consistently so as to minimize the total energy. Within this self-consistent approach, the total energy is expressed as the sum of the rotational energy and the potential energy associated with nuclear stretching.

Mariscotti {\it et al} ~\cite{Mariscotti69} demonstrated that stretching alone cannot account for the rapid increase in the moment of inertia, indicating that factors like Coriolis effects are essential. Due to the absence of detailed knowledge of the change in nuclear structure as a function of I at that time, Mariscotti proposed a phenomenological approach wherein the deformation variable was replaced by a generalized variable $t$ $\left(\Im\approx t^n\right)$, His analysis showed that the best fit for experimental data occurred when $n=1$ indicating a linear dependence of $\beta$, a result that is in clear contradiction to Bohr’s quadratic $\left(\Im\propto\beta^2\right)$ formula. The quantum mechanical formulation of the variable moment of inertia was latter established by several authors ~\cite{Das70,Volkov71}.

In this work, the problem of the moment of inertia is analyzed within the framework of the Cranking Model, considering only the collective motion of the nucleus, while microscopic single-particle dynamics are excluded. Following Diamond, we assume that the nucleus adjusts its shape self-consistently so as to minimize total energy. Within this framework, a distinction is made between two types of deformation: static deformation, which arises naturally from centrifugal stretching, and dynamic deformation, representing small vibrational deviations of the nuclear surface. Consequently, $\beta$ is treated as the primary dynamical variable, while additional physical factors are incorporated through the stiffness constants $C$ and initial moment of inertia $\Im_{0}$. By performing a Taylor expansion to a second order, we demonstrate that a linear dependence of the moment of inertia on $\beta$ serves as a mathematically sound approximation for small stretching. This analysis shows clearly that both the Bohr and Mariscotti models arise naturally as limiting cases of the Cranking Model.

\section{FORMALISM}

In the rotational (body-fixed) frame, pure rotational motion is absent, and only surface vibrations of the nucleus around its equilibrium deformed shape are observed. The components of vibrational angular momentum $\mu$ can be expressed in terms of the coupling between the spherical coordinates and their conjugate momenta as follows ~\cite{Greiner1969,Eisenberg 1970}:

\begin{eqnarray}
	L_{\mu'\mu''}^{vib}=i\sqrt{10}\sum_{\mu'}\left\langle 22\mu'\mu''| 1\mu\right\langle\left(a_{\mu'}^{0}+\hat{a}_{\mu'}\right)\pi_{\mu''}^{*}
	\label{eq:one}.
\end{eqnarray}

Here the deformation parameter is decomposed into two components: the static component $a_{\mu'}^{0}$ which represents the static equilibrium deformation at a specific spin  $I$. Physically, it corresponds to the stretching of the nucleus, due to the centrifugal force. Where $\hat{a}_{\mu'}$ is an operator represents the vibrational deviation from the static equilibrium, and $\pi_{\mu''}^{*}$ is the conjugate momentum of $\hat{a}_{\mu'}$. The $x$– component of the angular momentum is then given by

\begin{widetext}
		\begin{eqnarray}
		L_{x}&=&\frac{-1}{\sqrt{2}}\left(L_{+1}-L_{-1}\right) \nonumber \\
		&=&\frac{-1}{\sqrt{2}}i\sqrt{10}\sum_{\mu'}\left\langle 22\mu'\left(1-\mu'\right)| 11\right\rangle\left(a_{\mu'}^{0}+\hat{a}_{\mu'}\right)\pi_{\left(1-\mu'\right)}^{*} \nonumber \\
		&&-\left\langle 22\mu'\left(1-\mu'\right)| 1-1\right\rangle\left(a_{\mu'}^{0}+\hat{a}_{\mu'}\right)\pi_{\left(-1-\mu'\right)}^{*}  \nonumber \\
		&=&-i\sqrt{5}\left[\sum_{\mu'}\left\langle 22\mu'\left(1-\mu'\right)| 11\right\rangle a_{\mu'}^{0}\pi_{\left(1-\mu'\right)}^{*}-\left\langle 22\mu'\left(-1-\mu'\right)| 1-1\right\rangle a_{\mu'}^{0}\pi_{\left(-1-\mu'\right)}^{*}\right.   \nonumber \\
		&&\left. +\sum_{\mu'}\left\langle 22\mu'\left(1-\mu'\right)|11\right\rangle\hat{a}_{\mu'}^{0}\pi_{\left(1-\mu'\right)}^{*}-\left\langle 22\mu'\left(-1-\mu'\right)|1-1\right\rangle\hat{a}_{\mu'}\pi_{\left(-1-\mu'\right)}^{*}\right]. 
		\label{eq:wideeq}
\end{eqnarray}
\end{widetext}

where
\begin{eqnarray}
	\pi_{\mu''}^{*}&=&-i\sqrt{\frac{\hbar B_{2}\omega_{2}}{2}}\left[\left(-1\right)^{\mu''}b_{-\mu''}^{+}-b_{\mu''}\right] \nonumber\\
	\hat{a}_{\mu'}&=&-i\sqrt{\frac{\hbar}{2B_{2}\omega_{2}}}\left[b_{\mu'}^{+}+\left(-1\right)^{\mu'}b_{-\mu'}\right],	
\end{eqnarray}

\noindent are operators representing the displacement from the equilibrium deformation and its conjugate momentum, written in spherical coordinates, respectively. Here, $B_{2}$ denotes the quadrupole vibrational mass parameter associated with collective surface oscillations of the nucleus, and $\omega_{2}$ is the corresponding vibrational frequency.

The first term on the RHS of Eq. (2) represents transitions between two states $\left. |0\right\rangle $ and $\left. \beta_{K}^{\dagger}|0\right\rangle$ having the same projection of the angular momentum $K$ onto the intrinsic symmetry axis, but differ by one phonon. The second one corresponds to transitions between states $\left. \beta_{K}^{\dagger}|0\right\rangle$ and $\left. \beta_{K'}^{\dagger}|0\right\rangle$ with the same number of phonons but differing by one unit in $K$. by employing the first identity given in Eq. (3), the static part of Eq. (2) can be written as

\begin{widetext}
	\begin{eqnarray}
	L_{x}^{static}=&&-\frac{\sqrt{5}}{2}\sqrt{\hbar B_{2}\omega_{2}}
	\sum_{\mu'}\left[\left\langle 22\mu'\left(1-\mu'\right)|11\right\rangle\left( \left(-1\right)^{(1-\mu')}b_{-\left(1-\mu'\right)}^{+}+b_{\left(1-\mu'\right)}\right)\right. \nonumber\\ 
	&&-\left. \left\langle 22\mu'\left(-1-\mu'\right)|1-1\right\rangle\left( \left(-1\right)^{(1+\mu')}b_{\left(1+\mu'\right)}^{+}+b_{-\left(1+\mu'\right)}\right)\right]a_{\mu}^{0}.	
\end{eqnarray}

And the expectation value of $L_{x}^{static}$ is given by

	\begin{eqnarray}
	\left\langle 0\left|L_{x}^{static}b_{K}^{+} \right| 0\right\rangle 	=&&-\frac{\sqrt{5}}{2}\sqrt{\hbar B_{2}\omega_{2}}
		\sum_{\mu'}\left[\left\langle 22\mu'\left(1-\mu'\right)|11\right\rangle\left\langle 0\left| \left( \left(-1\right)^{(1-\mu')}b_{-\left(1-\mu'\right)}^{+}b_{K}^{+}+b_{\left(1-\mu'\right)}b_{K}^{+}\right)\right| 0\right\rangle \right. \nonumber\\ 
		&&-\left. \left\langle 22\mu'\left(-1-\mu'\right)|1-1\right\rangle\left\langle 0\left|\left( \left(-1\right)^{(-1-\mu')}b_{-\left(-1-\mu'\right)}^{+}b_{K}^{+}+b_{\left(1+\mu'\right)}b_{K}^{+}\right| 0\right\rangle \right) \right]a_{\mu'}^{0}.	
	\end{eqnarray}
\end{widetext}

The terms containing two-creation operators vanish when acting on the vacuum state. Using bosonic commutation relation $b_{\mu}b_{\mu'}^{+}=\delta_{\mu\mu'}+b_{\mu'}^{+}b_{\mu}$, Eq. (5) is simplifies to

\begin{widetext}
\begin{eqnarray}
	\left\langle 0\left|L_{x}^{static}b_{K}^{+}\right| 0\right\rangle=-\frac{\sqrt{5}}{2}\sqrt{\hbar B_{2}\omega_{2}}
	\left(\left\langle 22\left(1-K\right),K|11\right\rangle a_{1-K}^{0}-
	\left\langle 22\left(K-1\right)-K|1-1\right\rangle a_{K-1}^{0}\right) .	
\end{eqnarray}
\end{widetext}

In the case of axially symmetric nuclei where $a_{2\mu'}^{0}=\beta\delta_{\mu',0}$ the condition $\mu'=0$ is required. Consequently, the quantum number $K$ can take only the value $K=+1$. Substituting the matrix element of Eq. (6) into the Inglis formula ~\cite{Inglis 1954} and restricting the summation to $K=1$ we get the following

\begin{eqnarray}
	\Im^{static}&=&2\sum_{K}\frac{\left| \left\langle 0\left| L_{x}b_{K}^{+}\right| 0\right\rangle \right|^{2}}{\hbar\omega} \nonumber\\
	&=&\frac{5}{2}\hbar B_{2}\omega_{2}\beta^{2}
	\frac{\left| \left\langle 2201\left| 11\right. \right\rangle -\left\langle 220-1\left| 1-1\right. \right\rangle \right|^{2}}{\hbar\omega} \nonumber\\
	&=&3B\beta^{2}.		
\end{eqnarray}

This result clearly shows that, in the limiting case where the dynamical coupling between rotational and vibrational degrees of freedom is neglected, the Inglis formula reduces identically to the Bohr's expression for the moment of inertia.
  
\subsection{The dynamic part}

The dynamic term in Eq. (2) can be simplified as

\begin{eqnarray}
	L_{x}^{Dyn}=&&\frac{\hbar\sqrt{10}}{\sqrt{2}}\left\lbrace \sum_{\mu'}\left\langle 22\mu'\left( 1-\mu'\right)\left| 11\right. \right\rangle \left(-1\right)^{1-\mu'}b_{-\left(1-\mu'\right)}b_{\mu'}^{+}\right.   \nonumber\\
	&&\left. -
	\sum_{\mu'}\left\langle 22\mu'\left(-1-\mu'\right)\left| 1-1\right. \right\rangle \left(-1\right)^{1+\mu'}b_{\left(1+\mu'\right)}b_{\mu'}^{+}\right\rbrace.		
\end{eqnarray}

The matrix elements for this dynamic part involve transitions between two states $\left. \beta_{K}^{\dagger}|0\right\rangle$ and $\left. \beta_{K'}^{\dagger}|0\right\rangle$ having the same number of phonons but $K$ differs by $1$ as discussed previously. Accordingly one finds 

\begin{widetext}
	\begin{eqnarray}
		\left\langle 0\left|b_{K}L_{x}^{Dyn}b_{K'}^{+}\right|0\right\rangle 	=&&\frac{\hbar\sqrt{10}}{\sqrt{2}}\sum_{\mu'}\left(-1\right)^{\left(1-\mu'\right)}
		\left[\left\langle 22\mu'\left(1-\mu'\right)\left| 11\right.\right\rangle  
		\left\langle 0\left| b_{K}b_{-\left(1-\mu'\right)}b_{\mu'}^{+}b_{K'}^{+}\right| 0\right\rangle\right. \nonumber\\ 
		&&-\left.\left\langle 22\mu'\left(-1-\mu'\right)\left| 1-1\right.\right\rangle  
		\left\langle 0\left| b_{K}b_{\left(1+\mu'\right)}b_{\mu'}^{+}b_{K'}^{+}\right| 0\right\rangle\right] \nonumber\\ 
		=&&\frac{\hbar\sqrt{10}}{\sqrt{2}}\left(-1\right)^{\left(1-K\right)}\left(C_{1}\delta_{K',K-1}-C_{2}\delta_{K',K+1}\right),
	\end{eqnarray}
\end{widetext}

\noindent where we have used the commutation relations $b_{\mu}b_{\sigma}^{+}=\delta_{\mu\sigma}+b_{\sigma}^{+}b_{\mu}$ and we have dropped the terms including operators $b_{\mu}b_{\sigma}$ and $b_{\mu}^{+}b_{\sigma}^{+}$ to go from line 1 to line 2 in Eq. (9), here $C_{1}$ and $C_{2}$ are the Clebsh-Gordan coefficients,
where $C_{1}=\left\langle 22\left(1-K\right)K\left|11\right.\right\rangle$ and $C_{2}=\left\langle 22K\left(-1-K\right)\left|1-1\right.\right\rangle$.

Substitute this matrix element into the Inglis formula we obtain the corresponding contribution the the moment of inertia.

\begin{eqnarray}
	\Im^{Dyn}&=&2\sum_{K'}\frac{\left| \left\langle 0\left| b_{K}L_{x}b_{K}^{+}\right| 0\right\rangle \right|^{2}}{E_{K'}-E_{K}} \nonumber\\ &=&\frac{6\hbar}{E_{1}-E_{0}}=\eta.		
\end{eqnarray}

For practical purposes, the energy difference $E_1-E_0$ treated as a fitting parameter and the resulting contribution of $\Im^{Dyn}$ is denoted by $\eta$. As shown by the present analysis this term originates from the Coriolis effect. It follows that the total moment of inertia can be written as

\begin{eqnarray}
	\Im_{total}=3B\beta^{2}+\eta.		
\end{eqnarray}

Eq. (11) shows that the moment of inertia depends not only on the deformation parameter $\beta$, and the inertial parameter $B$, but also it depends on the Coriolis effect; The Coriolis term gives a strong contribution in the beta $(\beta)$ and gamma $(\gamma)$ bands,  while it is weak in the case of ground state band. This explains qualitatively why the moment of inertia in the case of beta and gamma bands differs from that of ground state bands.

\subsection{Variable moment of inertia equation}

The nucleus undergoes progressive stretching along its axis of symmetry with increasing angular momentum $I$, an additional contribution to the potential energy emerges. Consequently, the total energy of the system at a given angular momentum $I$ can be written as ~\cite{BohrMot75,Diamond64}: 

\begin{eqnarray}
	E=\frac{1}{2}C\left(\beta-\beta_{0}\right)^{2}+\frac{\hbar^{2}I\left(I+1\right) }{2\Im_{total}},		
\end{eqnarray}

\noindent where $\beta_{0}$ is the deformation parameter at zero angular momentum $(I=0)$. Using the definition of the moment of inertia defined in Eq. (11) and writing $\beta=\beta_{0}\left(1+\bigtriangleup\beta/\beta_{0}\right)$, Eq. (12) can be rewrite as:

\begin{eqnarray}
E=&&\frac{1}{2}C\left(\bigtriangleup\beta\right)^{2}+
\frac{\hbar^{2}I\left(I+1\right)}{2
		\left[3B\beta_{0}^{2}\left(1+\frac{\bigtriangleup\beta}{\beta_{0}}\right)^{2}+\eta\right]}, \nonumber\\ 	
=&&\frac{1}{2}C\left(\bigtriangleup\beta\right)^{2}+
\frac{\hbar^{2}I\left(I+1\right)}{
	6B\beta_{0}^{2}\left(1+\frac{2\bigtriangleup\beta}{\beta_{0}}+\frac{\eta}{3B\beta_{0}^{2}}\right)},		
\end{eqnarray}

A formula for the energy, closely resembling Eq. (13), was derived by the authors within the framework of the quantization of centrifugal stretching ~\cite{Mohd020,Mohd022}. 
The mathematical term $\left(1+\frac{2\bigtriangleup\beta}{\beta_{0}}\right)^{2}$ in the first line of Eq. (13) is approximated as $1+\frac{2\bigtriangleup\beta}{\beta_{0}}$ through a second-order Taylor expansion yielding the expression in the second line. Straightforward calculations then shows that the total energy in Eq. (13) for the spin $I$ can be written exactly in a form where $E$ is expressed in terms of the moment of inertia, the form that was suggested by Mariscotti et al ~\cite{Mariscotti69} as:

\begin{eqnarray}
	E=\frac{1}{2}C'\left(\Im-\Im_{0}\right)^{2}+\frac{\hbar^{2}I\left(I+1\right) }{2\Im},		
\end{eqnarray}

\noindent where
\begin{eqnarray}
	\Im_{0}&=&3B\beta_{0}^{2}+\eta \nonumber\\
	C'&=&\frac{C}{36B^{2}\beta_{0}^{2}}.	
\end{eqnarray}

According to Equation (15), the parameter $C'$ is invariant under Coriolis effects, while only the parameter $\Im_{0}$ depends on them.

In reference ~\cite{Okhunov023} the authors also present an alternative analysis, closely related to the present approach, demonstrating that the Bohr formula must be multiplied by an appropriate correction factor in order to reproduce the experimental values of $\Im_{0}$.

\section{CONCLUSIONS}

By applying the cranking model and restricting the analysis to collective motion, it is shown that both the Bohr and Variable Moment of Inertia (VMI) models arise naturally as limiting cases of the cranking framework. The present analysis clarifies the apparent contradiction between the Bohr and VMI descriptions by demonstrating that, in the regime of small centrifugal stretching, the quadratic dependence on the deformation parameter $\beta$ can be accurately approximated by a linear term through a second-order Taylor expansion. In addition, the contribution of the Coriolis effect is explicitly identified and quantified, in contrast to earlier treatments, such as that of Mariscotti et al., where the dependence was noted without a detailed exposition of its impact.

\vspace*{.5cm}



\nocite{*}
\bibliographystyle{elsarticle-num}

\end{document}